\documentclass{article}
\usepackage{ijcai21}
\usepackage{times}
\usepackage{soul}
\usepackage{url}
\usepackage{xcolor}
\usepackage[colorlinks=true,linkcolor=LightBlue,citecolor=DeepBlue,urlcolor=DeepBlue]{hyperref}
\usepackage[utf8]{inputenc}
\usepackage[small]{caption}
\usepackage{graphicx}
\usepackage{amsmath}
\usepackage{amsthm}
\usepackage{booktabs}
\usepackage{algorithm}
\usepackage[noend]{algorithmic}
\usepackage{multirow}
\usepackage[switch]{lineno}
\usepackage{amssymb}
\usepackage{amsbsy}
\usepackage{cleveref}
\usepackage{xspace}
\usepackage{subfig}
\usepackage{mdwlist}
\usepackage{enumitem}

\definecolor{DeepBlue}{rgb}{0.1,0.1,0.5}
\definecolor{LightBlue}{rgb}{0.1,0.1,0.8}
\setlist{nolistsep,leftmargin=*}

\newtheorem{thm}{\textbf{Theorem}}

\newtheorem{defn}{\textbf{Definition}}
\newtheorem{lem}{\textbf{Lemma}}

\newtheorem{cor}{\textbf{Corollary}}

\newcommand{\name}{\textsc{GraphReach}\xspace}
\newcommand{\gnn}{\textsc{Gnn}\xspace}
\newcommand{\gcn}{\textsc{Gcn}\xspace}
\newcommand{\gat}{\textsc{Gat}\xspace}
\newcommand{\gin}{\textsc{Gin}\xspace}
\newcommand{\gsage}{\textsc{GraphSage}\xspace}
\newcommand{\pgnn}{\textsc{P-Gnn}\xspace}
\newcommand{\CV}{\mathcal{V}\xspace}
\newcommand{\CE}{\mathcal{E}\xspace}
\newcommand{\CC}{\mathcal{C}\xspace}
\newcommand{\CX}{\boldsymbol{\mathcal{X}\xspace}}
\newcommand{\CW}{\mathcal{W}\xspace}
\newcommand{\CZ}{\mathbf{Z}\xspace}
\newcommand{\CS}{\mathcal{S}\xspace}
\newcommand{\CM}{\boldsymbol{\mathcal{M}\xspace}}
\newcommand{\CF}{\mathcal{F}\xspace}
\newcommand{\CA}{\mathcal{A}\xspace}
\newcommand{\cW}{\mathbf{W}\xspace}
\newcommand{\cx}{\mathbf{x}\xspace}
\newcommand{\ch}{\mathbf{h}\xspace}
\newcommand{\cz}{\mathbf{z}\xspace}
\newcommand{\rw}{\mathcal{R}\xspace}
\newcommand{\bg}{\mathcal{B}\xspace}

\DeclareMathOperator*{\argmax}{\arg\,\max}

\newcommand{\grid}{Grid\xspace}
\newcommand{\communities}{Communities\xspace}
\newcommand{\email}{Email\xspace}
\newcommand{\emailc}{Email-Complete\xspace}
\newcommand{\ppi}{PPI\xspace}
\newcommand{\protein}{Protein\xspace}
\newcommand{\cora}{CoRA\xspace}
\newcommand{\citeseer}{CiteSeer\xspace}

\newcommand\bld[2]{$\mathbf{#1 \pm #2}$}

\title{GraphReach: Position-Aware Graph Neural Network \\ using Reachability Estimations}

\author{
Sunil Nishad$^1$ \and
Shubhangi Agarwal$^1$ \and
Arnab Bhattacharya$^1$ \And
Sayan Ranu$^2$ \\
\affiliations
$^1$Indian Institute of Technology Kanpur, India \\
$^2$Indian Institute of Technology Delhi, India \\
\emails
snishad@cse.iitk.ac.in, sagarwal@cse.iitk.ac.in, arnabb@cse.iitk.ac.in,
sayanranu@cse.iitd.ac.in
}

\newcommand{\figwidth}{0.195\textwidth}

\begin{document}

\maketitle

\begin{abstract}
	Majority of the existing graph neural networks (\gnn) learn node embeddings
	that encode their local neighborhoods but not their positions.
	Consequently, two nodes that are vastly distant but located in similar
	local neighborhoods map to similar embeddings in those networks. This
	limitation prevents accurate performance in predictive tasks that rely on
	position information.  In this paper, we develop \name, a
	\emph{position-aware} inductive \gnn that captures the global positions of
	nodes through \emph{reachability estimations} with respect to a set of
	\emph{anchor} nodes. The anchors are strategically selected so that
	reachability estimations across all the nodes are maximized. We show that
	this combinatorial anchor selection problem is NP-hard and, consequently,
	develop a greedy $(1-1/e)$ approximation heuristic. Empirical evaluation
	against state-of-the-art \gnn architectures	reveal that \name provides up
	to $40\%$ relative improvement in accuracy. In addition, it is more robust
	to adversarial attacks.
\end{abstract}

\section{Introduction and Related Work}
\label{sec:intro}

Learning feature-space node embeddings through graph neural networks (\gnn) has
received much success in tasks such as node classification, link prediction,
graph generation, learning combinatorial algorithms, etc.
\cite{hamilton2017graphsage,kipf2017iclr,iclr2018gat,graphgen,gcomb}.  {\gnn}s
learn node embeddings by first collecting structural and attribute information
from the neighborhood of a node and then encoding this information into a
feature vector via non-linear transformation and aggregation functions.  This
allows {\gnn}s to generalize to unseen nodes, i.e., nodes that were not present
during the training phase.  This \emph{inductive} learning capability of
{\gnn}s is one of the key reasons behind their popularity.

\begin{figure}[t]
    \centering
    \subfloat[]{
    \label{fig:pgnn}
    \includegraphics[width=0.50\columnwidth]{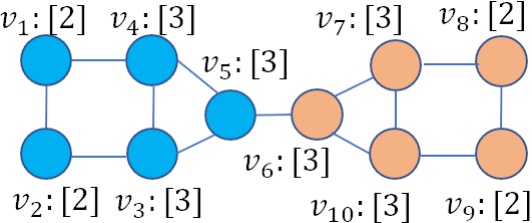}}
    \hfill
    \subfloat[]{
    \label{fig:motivation}
    \includegraphics[width=0.33\columnwidth]{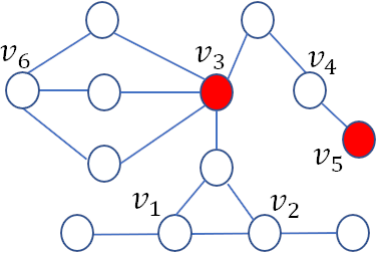}}
	\caption{{(a) The color of the node indicates its
	class label. Each node is also characterized by a numerical attribute. (b)
	The red nodes, $v_3$ and $v_5$, represent the anchor nodes.}}
    \label{fig:my_label}
\end{figure}

Due to reliance on only the neighborhood information, most \gnn architectures
fail to distinguish nodes that are located in different parts of the graph, but
have similar neighborhoods. In the extreme case, if two nodes are located in
topologically isomorphic neighborhoods, their learned embeddings are identical.
To elaborate, consider nodes $v_1$ and $v_8$ in Fig.~\ref{fig:pgnn}. The two
nodes belong to two different class labels (node color). However, since their
2-hop neighborhoods are isomorphic to each other, their learned embeddings in a
2-layer Graph Convolutional Network (\gcn) are identical, despite the fact that
they are far away in the graph. Hence, the \gcn will be incapable of correctly
predicting that $v_1$ and $v_8$ belong to different class labels. Consequently,
for predictive tasks that rely on position of a node with respect to the graph,
the performance suffers. 
While node embedding methods such as DeepWalk~\cite{perozzi2014deepwalk} and
Node2Vec~\cite{grover2016node2vec} do learn position information, they are
\emph{transductive} in nature, i.e., they fail to generalize to unseen nodes.
In addition, they cannot encode attribute information. Our goal, therefore, is
to design a \gnn that is inductive, captures both position as well as node
attribute information in the embeddings, and overcomes the problem of
automorphism leading to identical embeddings.

We emphasize that we do not claim position information to be more important
than the neighborhood structure. Rather, the more appropriate choice depends on
the problem being solved and the dataset. For example, if homophily is the main
driving factor behind a prediction task, then embeddings based on local
neighborhood structure are likely to perform well. On the other hand, to predict
whether two nodes belong to the same community, e.g., road networks, gene
interaction networks, etc., positional information is more important.

\pgnn \cite{you2019pgnn} is the first work to address the need for an inductive
\gnn that encodes position information. \pgnn randomly selects a small number
of nodes as \emph{anchor} nodes. It then computes the shortest path distances
of all remaining nodes to these anchor nodes, and embeds this information in a
low-dimensional space.  Thus, two nodes have similar embeddings if their
shortest path distances to the anchors are similar. To illustrate, we revisit
Fig.~\ref{fig:pgnn}, and assume that $v_7$ is an anchor node. Although the
2-hop neighborhoods of $v_1$ and $v_8$ are isomorphic, their shortest path
distances to $v_7$ are different and, hence, their embeddings will be different
as well.  Although \pgnn has shown impressive improvements over existing \gnn
architectures in link prediction and pairwise node classification, there is
scope to improve. 

\emph{1.~Holistic Position Inference:} Since \pgnn relies \emph{only} on
shortest path distances to compute embeddings, remaining paths of the graph are
ignored. Consequently, two nodes can be incorrectly concluded
to be in the same position, even if they are not. Fig.~\ref{fig:motivation}
illustrates this aspect.
Here, nodes $v_1$, $v_2$, and $v_6$ are all equidistant from the anchors $v_3$ and $v_5$ and
consequently, their embeddings would be similar. Note, however, that $v_6$ is
located in a different region with several paths of length $2$ to $v_3$,
whereas $v_1$ and $v_2$ have only one path. Thus, $v_6$ should have a different
embedding.

\emph{2.~Semantics:} The assumption of shortest path rarely holds in the real
world.  This has been shown across varied domains such as Wordmorph \cite{sp1},
Wikispeedia \cite{sp2}, metro transportation \cite{sp3}, flight connections
\cite{sp4}, etc.  Instead, \emph{many} short paths are followed~\cite{sp2}.

\emph{3.~Robustness to Adversarial Attacks:} Relying only on shortest paths
also makes \pgnn vulnerable to adversarial attacks. Specifically, adding a
small number of critical edges in the graph can significantly alter the
shortest path distances for targeted nodes and, hence, their node embeddings.

\begin{table}[t]
	\centering
\scalebox{0.75}{
	\begin{tabular}{llp{1.6in}}
		\toprule
		\textbf{Symbol} & \textbf{Dimensions} & \textbf{Description}\\
		\midrule
		$\CX=\{ \cx_v\mid\forall v\in\CV\}$ & $n \times d$ & Original node attributes\\
		$\mathbf{H}^l=\{ \ch^l_v\mid\forall v\in\CV\}$ & $n \times d_{hid}$ & Feature matrix at layer $l$\\
		$\CM^l=\{ \CM^l_v\mid\forall v\in\CV\}$ & $n \times k \times d_{hid}$ & Message Tensor at layer $l$\\
		$\cW^l_{\CM}$ &  $2 \cdot d_{hid} \times d_{hid}$ & Transform $\CM^l$\\
		$\mathbf{a}^l_{att}$ & $2 \cdot d_{hid} \times 1$ & Attention vector at layer $l$\\
		$\CZ=\{ \cz_v\mid\forall v\in\CV\}$ & $n \times k$ & Output embeddings\\
		$\cW_Z$ & $d_{hid} \times 1$ & Transforms $\CM^L$ to $\mathbf{Z}$\\
		\bottomrule
	\end{tabular}}
	\caption{Matrix notations and descriptions.}
	\label{tab:matrices}
\end{table}

To overcome these shortcomings, in this paper, we introduce a new
position-aware \gnn architecture called \textbf{\name}\footnote{Appendices 
are in \url{https://arxiv.org/abs/2008.09657/}.}. Our key contributions are
as follows:

\begin{itemize}

	\item \name computes position-aware inductive node embeddings through
		\emph{reachability estimations}.  Reachability estimations encode the
		position of a node with respect to \emph{all} paths in the graph
		(\S\ref{sec:formulation} and \S\ref{sec:graphreach}).  This
		\emph{holistic} position estimation also ensures that \name is robust
		to adversarial attacks since a few edge additions or deletions do not
		substantially alter the reachability likelihoods
		(\S\ref{sec:adversarial}).

	\item Unlike \pgnn, where anchors are randomly selected, \name
		strategically selects anchors by framing it as a problem of maximizing
		\emph{reachability} along with their \emph{attention} weights. We show
		that maximizing reachability is NP-hard, monotone and submodular. We
		overcome this bottleneck through a greedy hill-climbing algorithm,
		which provides a $(1-1/e)$ approximation guarantee (\S\ref{sec:anchor}).

	\item Extensive empirical evaluation across $8$ datasets demonstrates a
		dramatic relative improvement of up to $40\%$ over \pgnn and other
		state-of-the-art \gnn architectures (\S\ref{sec:expts}).

\end{itemize}

\section{Problem Formulation}
\label{sec:formulation}

We represent a graph as $G=(\CV,\CE,\CX,\CW)$, where $\CV$ is the set of nodes
$v_i$, with $1 \leq i \leq n$ and $\CE$ is the set of edges. The attribute set
$\CX=\{\cx_{v_1}, \cdots, \cx_{v_n}\}$ has a one-to-one correspondence with the
node set $\CV$ where $\cx_{v_i}$ represents the feature vector of node
$v_i\in\CV$. Similarly, $\CW$ has a one-to-one correspondence with $\CE$, where
$w_{e_i}$ denotes the edge weights.

A \emph{node embedding model} is a function $f:\CV\rightarrow \CZ$ that maps
the node set $\CV$ to a $d$-dimensional vector space
$\CZ=\{\cz_1,\cdots,\cz_n\}$.  Our goal is to learn a \emph{position-aware}
node embedding model \cite{you2019pgnn}.

\begin{defn}[Position-aware Embedding]
	A node embedding $\cz_{v},\forall v \in \CV$ is position-aware if there
	exists a function $g(\cdot,\cdot)$ such that $d(v_i,v_j) = g(\cz_i,\cz_j)$,
	where $d(v_i,v_j)$ is the distance from $v_i$ to $v_j$ in $G$.
\end{defn}

The distance $d(v_i,v_j)$ should reflect the quality of all paths between $v_i$
and $v_j$ wherein, {(1)}~$d(v_i,v_j)$ is directly proportional to the number of
paths between $v_i$ and $v_j$, and {(2)}~$d(v_i,v_j)$ is inversely proportional
to the lengths of the paths between $v_i$ and $v_j$.  We capture these aspects
in the form of \emph{reachability estimations} through \emph{random walks}.
Note that, $d(\cdot,\cdot)$ is not required to be a metric distance function.

Reachability estimations are similar to PageRank \cite{pagerank}
and Random Walk with Restarts \cite{rwr}.  To the best of our knowledge, GIL
\cite{gil} is the only other \gnn framework that uses the concept of reachability
estimations. However, GIL does not use reachability to learn node embeddings.
Rather, they are used as additional features along with node embeddings for
node classification. 

\noindent
\textbf{Reachability Estimations:}
In a fixed-length random walk of length $l_w$, we start from an initial node
$v_i$, and jump to a neighboring node $u$ through an outgoing edge $e=(v_i,u)$
with \emph{transition probability}
$p(e) = w_{e} / \sum_{\forall e'\in N(v_i)} w_{e'}$.
Here, $N(v_i)$ denotes the set of outgoing edges from $v_i$. From node $u$, the
process continues iteratively in the same manner till $l_w$ jumps.  If there are
many short paths from $v_i$ to $v_j$, there is a high likelihood of reaching $v_j$
by a \emph{random walk} from $v_i$. We utilize this property to define a similarity
measure:
{\small
\begin{align}
\label{eq:sim}
s(v_i,v_j)=\frac{\sum_{k=1}^{n_w} count_{k}(v_i, v_j)}{l_w\times n_w}
\end{align}
}
Here, $n_w$ denotes the number of random walks started from $v_i$, and
$count_{k}(v_i, v_j)$ denotes the number of times random walker visited
$v_j$ in the $k^\text{th}$ random walk starting from $v_i$. The similarity
function could also weight the nodes according to the order they appear 
in a random walk (refer to App.~A). From $s(v_i,v_j)$, 
one may define $d(v_i,v_j)=1-s(v_i,v_j)$. However, we directly
work with similarity $s(\cdot,\cdot)$ since $d(v_i,v_j)$ is not explicitly
required in our formulation.

\begin{algorithm}[t]
	\caption{\name }
	\label{algo:pseudocode}
	{\footnotesize
	\textbf{Input:} Graph $G=(\CV,\CE,\CX,\CW)$; Anchors $\{a_i\}$;
	Message computation function $\mathcal{F}$; Message aggregation function $\CS$;
	Number of layers $L$; Non-linear function $\sigma$ \\
	\textbf{Output:} Node embedding $\cz_v, \forall v \in \CV$ 
	\begin{algorithmic}[1]
		\STATE $\ch^0_v\gets \cx_v, \forall v \in \CV$
		\FOR {$l=1,\cdots,L$}
			\FOR {$v\in\CV$}	
				\FOR {$i=1,\cdots,k$}
					\STATE $\widehat{\CM}^l_v[i] \gets \CF(v,a_i,\ch^{l-1}_v,\ch^{l-1}_{a_i})$ $\rhd$ Msg Computation
				\ENDFOR
				\STATE $\CM^l_v = (\mathop{\oplus}_{a_i \in \mathcal{A}} \widehat{\CM}^l_v[i]) \cdot \cW^l_{\CM}$ $\rhd$ Concatenation: Eq.~\eqref{eq:msgtensor}
				\STATE $\ch^{l}_v\gets \CS\left(\CM^l_v\right)$ $\rhd$ Msg Aggr: Eq.~\eqref{eq:meanpool} and Eq.~\eqref{eq:attention}
			\ENDFOR
		\ENDFOR
		\RETURN $\cz_v \in \mathbb{R}^k \gets \sigma(\CM^L_v . \cW_Z), \forall v \in \CV$
	\end{algorithmic}
	}
\end{algorithm}

\section{\name}
\label{sec:graphreach}

The symbols and notations used through out this paper are summarized in Table
\ref{tab:matrices}. All the matrix and vector notations are denoted in bold
letters by convention.

\subsection{The Architecture}
\label{sec:arch}

\begin{figure}[t]
  \centering
  	\includegraphics[width=\columnwidth]{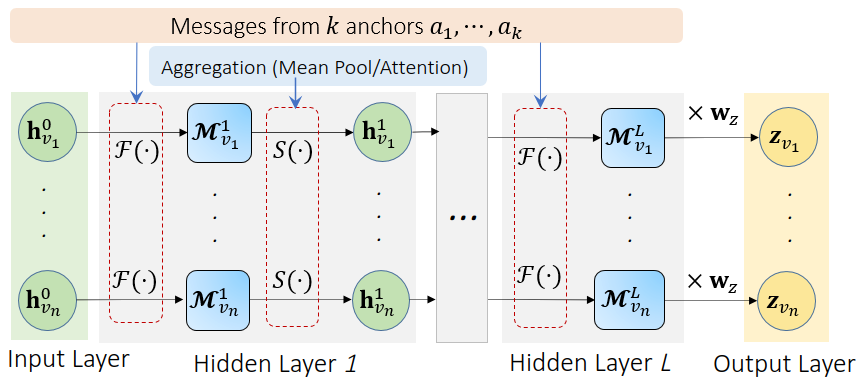}
	\caption{{The architecture of \name. Note that aggregation is not performed in the last hidden layer.}}
	\label{fig:arch}
\end{figure}

Algo.~\ref{algo:pseudocode} outlines the pseudocode and Fig.~\ref{fig:arch}
pictorially depicts the architecture. In addition to hyper-parameters and message
passing functions, the input to Algo.~\ref{algo:pseudocode} includes the
graph and $k$ anchor nodes.
The details of the anchor selection procedure are discussed in \S\ref{sec:anchor}.
In the initial layer, the embedding of a node $v$ is simply its attribute
vector $\cx_v$ (line 1). In each hidden layer, a set of \emph{messages},
$\CM^l$, is computed using the message computing function
$\CF(v,a,\ch_v^l,\ch_a^l)$ between each node-anchor pair (details in
\S\ref{subsec:msg_computation}) (lines 2-6).  The component $\CM^l_v[i]$
in $\CM^l$ represents the message received by node $v$ from the $i^{th}$
anchor node.
These messages are then aggregated using an aggregation function $\CS$ (line
7), which we will discuss in detail in \S\ref{subsec:msg_agg}. The aggregated
messages thus obtained are propagated to the next layer. In the final layer,
the set of messages for each node, i.e., $\CM^l_v$, is linearly
transformed through a trainable weight matrix $\cW_Z$ (line 8).

\subsection{Anchor Selection}
\label{sec:anchor}

Anchors act as our reference points while encoding node positions. It is
therefore imperative to select them carefully. Selecting two anchors that are
close to each other in the graph is not meaningful since the distance to these
anchors from the rest of the nodes would be similar.  Ideally, the anchors
should be diverse as well as \emph{reachable} from a large portion of nodes.

Formally, let $\rw$ be the set of all random walks performed across all
nodes in $\CV$. We represent the reachability information in the form of a
\emph{bipartite} graph $\bg = (\CV_1,\CV_2,\CE_B)$. Here, $\CV_1=\CV_2=\CV$.
There is an edge $e=(u,v)\in \CE_B,\;u\in\CV_1,\;v\in\CV_2$ if there exists a
walk in $\rw$ that starts from $v$ and \emph{reaches} $u$.  The
\emph{reachability set} of a subset of nodes $\CA$ is:
{\small
\begin{align}
\rho(\CA)=\{v\mid(u,v)\in \CE_B,\:u \in \CA, v\in\CV_2\} 
\end{align}
}

Our objective is to find the set of $k$ anchors $\CA^*\subseteq \CV_1$,
$|\CA^*|=k$, that \emph{maximizes} reachability. Specifically,
{\small
\begin{align}
\label{eq:reachability}
\CA^*=\argmax_{\CA\subseteq\CV_1, |\CA|=k}\{|\rho(\CA)|\}
\end{align}
}

\begin{lem}
The maximization problem in Eq.~\ref{eq:reachability}, performed on the bipartite graph formed on the reachability set, is NP-hard.
\end{lem}

\begin{lem}
	\label{lem:submodular}
	\textit{For any given set of nodes $\CA$, $f(\CA)=|\rho(\CA)|$ is monotone and submodular.}
\end{lem}

The proofs are in App.~B and App.~C respectively.

For \emph{monotone} and \emph{submodular} optimization functions, the
greedy-hill climbing algorithm provides a $(1-1/e)$ approximation
guarantee \cite{submodular}. We, thus, follow the same strategy and iteratively add the node that provides highest \emph{marginal} reachability (Algo.~2 in App.~D).

\begin{cor}
	\label{lem:guarantee}
	\textit{If set $\CA$ is the output of Algo.~2 (in App.~D), then $\mid\rho(\CA)\mid\geq \left(1-1/e\right)\mid\CA^*\mid$, where $\CA^*$ is the anchor set of size $k$ that maximizes reachability coverage.}
\end{cor}
\textsc{Proof.} Follows from Lemma~\ref{lem:submodular}.

\noindent
\textbf{Modeling Reachability Frequency:} The above approach models
reachability as a binary occurrence; even if there exists just one walk
where $v\in\CV_2$ reaches $u\in\CV_1$, an edge $(u,v)$ is present in $\bg$.
It does not incorporate the frequency with which $u$ is visited from $v$.
To capture this aspect, we randomly sample $X\%$ of the walks from $\rw$ 
and form the bipartite graph only on this sampled set. Note that the
down-sampling does not require the bipartite graph to be fully connected.
Algo.~2 (in App.~D) is next run to select anchors
on this bipartite graph. This process is repeated multiple times by drawing
multiple samples of the same size from $\rw$ and the final anchor set consists
of nodes that are selected in the answer sets the highest number of times.
In our experiments, we sample $5$ subsets with $X=30\%$.

\subsection{Message Computation}
\label{subsec:msg_computation}

The message computation function $\CF\left(v,a,\ch_v^l,\ch_a^l\right)$
should incorporate both the position information of node $v$ with respect to
the anchor set $\CA$ as well as the node attributes. While node attributes may
provide important information that may be useful in the eventual prediction
task, position information, in the form of reachability estimations, captures the
location of the node in the global context of the graph. To encode these dual
needs, $\CF\left(v,a,\ch_v^l,\ch_a^l\right)$ is defined as follows.
{\small
\begin{align} 
\label{eqn:msg_fn} 
\mathcal{F}\left(v, a,\ch^l_v, \ch^l_a\right) = \Big( \big(s(v,a) \times \ch^l_v\big) \parallel \big(s(a,v) \times \ch^l_a \big)\Big) 
\end{align}
}
where $\parallel$ denotes concatenation of vectors. The function $\mathcal{F}$
takes as input a node $v$, an anchor $a$ and their respective layer
attributes, $\ch^l_v$ and $\ch^l_a$. It returns a message vector, which is a
weighted aggregation of their individual attributes in proportion to their
reachability estimations (Eq.~\ref{eq:sim}). Observe that, the reachability
estimations are used in both directions to account for an asymmetric distance function.

Due to concatenation, the output message vector is of dimension $2 \cdot d_{hid}$, where $d_{hid}$ is dimension of the node embeddings in the hidden layers.
To linearly transform the message vector back into $\mathbb{R}^{d_{hid}}$, we multiply it with a weight vector $\cW^l_{\CM}$.
The complete global structure information for node $v$ is encompassed in
the message matrix $\CM^l_v$ ($\oplus$ denotes row-wise stacking of message vectors). 
{\small
\begin{align}
	\label{eq:msgtensor}
	\CM^l_v = \Big(\mathop{\oplus}_{a \in \mathcal{A}} \mathcal{F}\left(v, a, \ch^l_v, \ch^l_a\right)\Big) \cdot \cW^l_{\CM}
\end{align}
}
%

\subsection{Message Aggregation}
\label{subsec:msg_agg}

To compute the hidden representation of nodes, messages corresponding to
anchors are aggregated for each node. We propose two aggregation schemes.

 \textbf{1. Mean Pool ($M$)}: In this, a simple mean of the message vectors
		are taken across anchors.
		{\small
		\begin{align}
			\label{eq:meanpool}
			\CS^M(\CM^l_v) = \frac{1}{k} \sum_{i=1}^k \CM^l_v[i]
		\end{align}
		}

\textbf{2. Attention Aggregator ($A$)}: In mean pooling, all anchors are
		given equal weight. Theorizing that the information being preserved can
		be enhanced by capturing the significance of an anchor with respect to
		a node, we propose to calculate the significance distribution among
		anchors for each node.  Following the Graph Attention Network (\gat)
		architecture \cite{iclr2018gat}, we compute attention coefficients of
		anchors for an anchor-based aggregation. The attention coefficient of
		the $i^\text{th}$ anchor $a_i$ is computed with
		trainable weight vector $\mathbf{a}^l_{att}$
		and weight matrix $\cW^l_{att}$.
		For node $v$, the attention weight with respect to anchor $i$ is
		{\small
		\begin{align}
			\alpha_v[i] = \text{sm} \bigg( \sigma_{att}\Big((\ch^l_v \cdot \cW^l_{att}  \parallel \CM^l_v[i] \cdot \cW^l_{att} ) \cdot \mathbf{a}^l_{att}  \Big) \bigg)
		\end{align}
		}
		Here, $sm$ denotes the softmax function.
		As followed in \gat architecture \cite{iclr2018gat}, we use \emph{LeakyReLU} as
		the non-linear function $\sigma_{att}$ (with negative input slope $0.2$).
		
		Finally, the messages are aggregated across anchors using these
		coefficients.
		{\small
		\begin{align}
			\label{eq:attention}
			\CS^{A}(\CM^l_v) = \sum_{i=1}^k \alpha_v[i] \times \CM^l_v[i] \cdot \cW^l_{att} + \ch^l_v \cdot \cW^l_{att} 
		\end{align}
		}

\subsection{Hyper-parameters for Reachability Estimation}
\label{sec:guarantees}

Reachability information relies on two key random walk parameters: the
length $l_w$ of each walk, and the total number of walks $n_w$. If $l_w$
is too short, then we do not gather information with respect to anchors
more than $l_w$-hops away. With $n_w$, we allow the walker to sample
enough number of paths so that our reachability estimations are accurate
and holistic. We borrow an important theorem from \cite{arrival} to guide
our choice of $l_w$ and $n_w$.
\begin{thm}
\cite{arrival} If there exists a path between two nodes $u$ and $v$ in a graph, with $1 - 1/n$ probability the random walker will find the path 
if the number of random walks conducted, $n_w$, is set to
$\Theta(\sqrt[3]{n^2 \ln n})$ with the length of each random walk, $l_w$,
being set to the diameter of the graph.
\end{thm}  

\subsection{Complexity Analysis}
\label{subsec:complexity}

We conduct $n_w$ random walks of length $l_w$ for all the $n$ nodes of the graph;
this requires $O(n_w \cdot l_w \cdot n)$ time. For anchor set selection, we sample
random walks multiple times and select $k$ anchors using the resulting bipartite
graphs formed. The complexity of sampling walks is $O(n \cdot n_w)$ while selecting
the anchors takes $O(k + k \cdot \log{k}) = O(k \cdot \log{k})$ operations.
Considering that each node communicates with $k$ anchors, there are ${O(n \cdot k)}$ message
computations. The aggregation of messages also requires $O(n \cdot k)$ operations
owing to $k$ messages being aggregated for each of the $n$ nodes.
The attention aggregator
has an additional step devoted to computation of attention coefficients which takes
$O(n \cdot k)$ time as well.

\section{Experiments}
\label{sec:expts}

In this section, we benchmark \name and establish that \name
provides up to $40\%$ relative improvement over state-of-the-art
\gnn architectures. The implementation is available at
\url{https://github.com/idea-iitd/GraphReach}.

\subsection{Experimental Setup}
\label{subsec:setup}
Please refer to App.~E for information on reproducibility.

\noindent
\textbf{Datasets:} We evaluate \name on the datasets listed in Table~\ref{tab:data_stat}.
Further details are available in App.~E.

\begin{table}[t]
\centering
\resizebox{\columnwidth}{!}{		
\begin{tabular}{lrrrrr}
		\toprule
		\textbf{Datasets} & \textbf{\#Nodes} & \textbf{\#Edges} & \textbf{\#Labels} & \textbf{Diameter} & \textbf{\#Attributes} \\
		\midrule
		\grid & 400 & 760 & - & 38 & -\\
		\communities & 400 & 3.8K & 20 & 9 & -\\
		\ppi & 56.6K& 818K & - & 8 & 50\\
		\emailc & 986 & 16.6K & 42 & 7 & -\\
		\email & 920 & 7.8K & 6 & 7 & -\\
		\protein & 43.4K & 81K & 3 & 64 & 29\\
		\cora & 2.7K & 5.4K & 7 & 19 & 1433\\
		\citeseer & 3.3K & 4.7K & 6 & 28 & 3703\\
		\bottomrule	
	\end{tabular}
}
	\caption{Characteristics of graph datasets used.}
	\label{tab:data_stat}
\end{table}

\noindent
\textbf{Baselines:} We measure and compare the performance of \name with five baselines:
{(1)} \pgnn~\cite{you2019pgnn}, {(2)} \gcn~\cite{kipf2017iclr},
{(3)} \gsage~\cite{hamilton2017graphsage}, {(4)} \gat~\cite{iclr2018gat}, and
{(5)} \gin~\cite{xu2019gin}.

\addtolength{\tabcolsep}{-3pt}

\begin{table*}[t]
\centering
\resizebox{\textwidth}{!}{
	\subfloat[{\large Pairwise Node Classification (PNC)}]{
\label{tab:pairwise}
\begin{tabular}{lcccc}\\ \toprule
	\textbf{Models} & \textbf{\communities} & \textbf{\email} & \textbf{\emailc} & \textbf{\protein} \\ \midrule
	\gcn & ${0.520 \pm 0.025}$ & ${0.515 \pm 0.019}$ & ${0.536 \pm 0.006}$ & ${0.515 \pm 0.002}$ \\
	\gsage & ${0.514 \pm 0.028}$ & ${0.511 \pm 0.016}$ & ${0.508 \pm 0.004}$ & ${0.520 \pm 0.003}$ \\
	\gat & ${0.620 \pm 0.022}$ & ${0.502 \pm 0.015}$ & ${0.511 \pm 0.008}$ & ${0.528 \pm 0.011}$ \\
	\gin & ${0.620 \pm 0.102}$ & ${0.545 \pm 0.012}$ & ${0.544 \pm 0.010}$ & ${0.523 \pm 0.002}$ \\
	\pgnn-F & ${0.997 \pm 0.006}$ & ${0.640 \pm 0.037}$ & ${0.630 \pm 0.031}$ & ${0.729 \pm 0.176}$ \\
	\pgnn-E & \bld{1.000}{0.001} & ${0.640 \pm 0.029}$ & ${0.637 \pm 0.037}$ & ${0.631 \pm 0.175}$ \\ 
	\midrule
	\name & \bld{1.000}{0.000} & $\mathbf{0.949 \pm 0.009}$ & $\mathbf{0.935 \pm 0.006}$ & $\mathbf{0.904 \pm 0.003}$\\ \bottomrule
\end{tabular}}
	\subfloat[{\large Link Prediction (LP)}]{
\label{tab:link}
\begin{tabular}{lccccc}\\ \toprule
	\textbf{Models} & \textbf{\grid-T} & \textbf{\communities-T} & \textbf{\grid} &
		\textbf{\communities} & \textbf{\ppi}                                      \\ \midrule
	\gcn & ${0.698 \pm 0.051}$ & ${0.981 \pm 0.004}$ & ${0.456 \pm 0.037}$ & ${0.512 \pm 0.008}$ & ${0.769 \pm 0.002}$ \\
	\gsage & ${0.682 \pm 0.050}$ & ${0.978 \pm 0.003}$ & ${0.532 \pm 0.050}$ & ${0.516 \pm 0.010}$ & ${0.803 \pm 0.005}$ \\
	\gat & ${0.704 \pm 0.050}$ & ${0.980 \pm 0.005}$ & ${0.566 \pm 0.052}$ & ${0.618 \pm 0.025}$ & ${0.783 \pm 0.004}$ \\ 
	\gin & ${0.732 \pm 0.050}$ & ${0.984 \pm 0.005}$ & ${0.499 \pm 0.054}$ & ${0.692 \pm 0.049}$ & ${0.782 \pm 0.010}$ \\
	\pgnn-F & ${0.637 \pm 0.078}$ & ${0.989 \pm 0.003}$ & ${0.694 \pm 0.066}$ & $\mathbf{0.991 \pm 0.003}$ & ${0.805 \pm 0.003}$ \\
	\pgnn-E & ${0.834 \pm 0.099}$ & ${0.988 \pm 0.003}$ & ${0.940 \pm 0.027}$ & ${0.985 \pm 0.008}$ & ${0.808 \pm 0.003}$ \\ 
	\midrule
	\name          & $\mathbf{0.945 \pm 0.021}$ & $\mathbf{0.990 \pm 0.005}$ & \bld{0.956}{0.014} & $\mathbf{0.991 \pm 0.003}$ & $\mathbf{0.810 \pm 0.002}$ \\ \bottomrule
\end{tabular}}
}
	\caption{ROC AUC. (Grid-T and Communities-T indicate performance in transductive settings.
\pgnn-E uses exact shortest path distance to all anchors while \pgnn-F is a fast variant of \pgnn that uses truncated 2-hop shortest path distance.)}
\end{table*}

\addtolength{\tabcolsep}{3pt}

\noindent
\textbf{Prediction Tasks:} We evaluate \name on the prediction tasks
of \emph{Link Prediction (LP)} and \emph{Pairwise node classification (PNC)}
using the \emph{Binary Cross Entropy} (BCE) loss with \emph{logistic
activation}, and on \emph{Node Classification (NC)} using the
\emph{Negative Log Likelihood} (NLL) loss.

\noindent
\textbf{Setting:}
For LP, we evaluate in both inductive and transductive settings,
whereas for PNC	, only inductive setting is used.

\noindent
\textbf{Default Parameters and Design Choices:} Unless specifically mentioned, we
set the number of anchors ($k$) as $\log^2 n$. While our experiments reveal that
a smaller number of anchors is sufficient, since \pgnn uses $\log^2 n$ anchors, we
keep it the same. The length of each random walk, $l_w$, is set to $graph\; diameter$,
and the number of walks $n_w$ as $50$. We also conducted experiments to analyze how
these parameters influence prediction accuracy of \name.

We use \emph{Attention} aggregation (Eq.~\ref{eq:attention}) and simple random
walk counts as the similarity function (Eq.~\ref{eq:sim}) to compare with the
baselines. For a fair comparison, values of parameters that are common to \name and other
baselines are kept the same. Their exact values are reported in App.~E.

\subsection{Comparison with Baselines}
\label{sec:results}

\noindent
\textbf{Pairwise Node Classification (PNC):} Table~\ref{tab:pairwise}
summarizes the performances in PNC. We observe a dramatic performance
improvement by \name over all existing \gnn architectures.  While \pgnn clearly
established that encoding global positions of nodes helps in PNC, \name further
highlights the need to go beyond shortest paths. Except in Communities, the
highest accuracy achieved by any of the baselines is $0.73$. In sharp contrast,
\name pushes the ROC AUC above $0.90$, which is a significant $\approx 40\%$
relative improvement, on average, over the state-of-the-art.  

\noindent
\textbf{Link Prediction (LP):} Table~\ref{tab:link} presents the results for LP.
Consistent with the trend observed in PNC, \name outperforms other {\gnn}s across
inductive and transductive settings. \pgnn and \name are significantly better than
the rest of the architectures. This clearly indicates that position-aware node
embeddings help. \name being better than \pgnn suggests that a holistic approach of
encoding position with respect to all paths is necessary.

The second observation from Table~\ref{tab:link} is that the performance of
position-\emph{unaware} architectures are noticeably better in transductive
setting. Since transductive setting allows unique identification of nodes through
one-hot encodings, traditional \gnn architectures are able to extract some amount
of position information, which helps in the prediction. In contrast, for both
\pgnn and \name, one-hot encodings do not impart any noticeable advantage as
position information is captured through distances to anchors.

We also compare the time taken by the two best performing models (in
App.~F). On average, \name is $2.5$ times faster than \pgnn
owing to its strategic anchor selection.

\addtolength{\tabcolsep}{-3pt}

\begin{table}[t]
	\centering
	\resizebox{\columnwidth}{!}{
	\begin{tabular}{ll cc cc cc cc cc cc}
	\toprule
		\multirow{2}{*}{\textbf{Task}}&\multirow{2}{*}{\textbf{Dataset}} & \multicolumn{2}{c}{\textbf{\gcn}} & \multicolumn{2}{c}{\textbf{\gsage}} & \multicolumn{2}{c}{\textbf{\gat}}& \multicolumn{2}{c}{\textbf{\gin}} & \multicolumn{2}{c}{\textbf{\pgnn}} & \multicolumn{2}{c}{\textbf{\name}}  \\ 
		&& \textbf{S+T}&\textbf{S}& \textbf{S+T}&\textbf{S}& \textbf{S+T}&\textbf{S}& \textbf{S+T}&\textbf{S}& \textbf{S+T}&\textbf{S}& \textbf{S+T}&\textbf{S}\\
		\midrule
		LP & \cora & $\textbf{0.86}$ & $0.59$ & $0.85$ & $0.53$ & $\textbf{0.86}$ & $0.51$ & $\textbf{0.86}$ & $0.59$ & $0.81$ & $0.77$ & $0.83$ & $\textbf{0.84}$ \\
		LP & \citeseer & $\textbf{0.87}$ & $0.61$ & $0.85$ & $0.56$ & $\textbf{0.87}$ & $0.56$ & $0.86$ & $0.68$ & $0.77$ & $\textbf{0.76}$ & $0.77$ & $0.74$ \\
		\midrule
		PNC & \cora & $\textbf{0.98}$& $0.50$ & $0.96$& $0.50$ & $\textbf{0.98}$&$0.51$ & $\textbf{0.98}$&$0.52$ & $0.86$&$0.59$ & $0.96$&$\textbf{0.77}$ \\
		PNC & \citeseer & $\textbf{0.96}$ & $0.51$ & $0.95$ & $0.51$ & $\textbf{0.96}$ & $0.50$ & $\textbf{0.96}$ & $0.53$ & $0.77$ & $0.57$ & $0.91$ & $\textbf{0.61}$ \\
		\midrule
		NC & \cora & $\textbf{0.92}$ & $0.52$ & $\textbf{0.92}$ & $0.50$ & $0.90$ & $0.50$ & $0.90$ & $0.54$ & $0.73$ & $0.50$ & $0.84$ & $\textbf{0.86}$ \\
		NC & \citeseer & $\textbf{0.82}$ & $0.52$ & $\textbf{0.82}$ & $0.50$ & $0.81$ & $0.50$ & $\textbf{0.82}$ & $0.53$ & $0.73$ & $0.55$ & $0.75$& $\textbf{0.71}$ \\
		\bottomrule
	\end{tabular}}
	\caption{ROC AUC. (`S' denotes the version containing only the graph structure and `S+T' denotes structure with  node attributes.)}
	\label{tab:cora_citeseer}
\end{table}

\addtolength{\tabcolsep}{3pt}

\subsection{Difference from Neighborhood-based {\gnn}s}

We conducted experiments on attributed graph datasets, with and without
attributes for prediction tasks. Table~\ref{tab:cora_citeseer} presents
the results for \cora and \citeseer on LP, PNC and NC.

In addition to the  network structures, both \cora and \citeseer,
are also accompanied by binary word vectors characterizing each node.
When the word vectors are ignored, the performance of neighborhood-aggregation
based {\gnn}s are significantly inferior ($\approx 25\%$) to \name.
When supplemented with word vectors, they outperform \name ($\approx
10\%$ better). This leads to the following conclusions. (1) Position-aware
{\gnn}s are better in utilizing the structural information. (2)
Neighborhood-aggregation based {\gnn}s may be better in exploiting
the feature distribution in the neighborhood. (This is not always though,
e.g., in \ppi and \protein, \name is better than the baselines even when
attribute information is used as seen in Tables~\ref{tab:pairwise} and
\ref{tab:link}). (3) The two approaches are \emph{complementary} in
nature and, therefore, good candidates for ensemble learning. To
elaborate, neighborhood aggregation based architectures rely on
\emph{homophily} \cite{homophily}. Consequently, if the node property is a
function of neighborhood attribute distribution, then neighborhood aggregation
performs well. On the other hand, in LP, even if the local neighborhoods of two
distant nodes are isomorphic, this may not enhance their chance of having a link.
Rather, the likelihood increases if two nodes have many neighbors in common.
When two nodes have common neighbors, their distances to the anchors are also
similar, and this positional information leads to improved performance.

\addtolength{\tabcolsep}{-4pt}

\begin{table*}[t]
\centering
\resizebox{\textwidth}{!}
{
	\subfloat[{\large Pairwise Node Classification (PNC)}]{
\label{tab:pairwise_ablation}
\begin{tabular}{lcccc}\\ \toprule
	\textbf{Models} & \textbf{\communities} & \textbf{\email} & \textbf{\emailc} & \textbf{\protein} \\
	\midrule
	GR-A     & \bld{1.000}{0.000} & $\mathbf{0.949 \pm 0.009}$ & ${0.935 \pm 0.006}$ & ${0.904 \pm 0.003}$\\ 
	GR-M     & \bld{1.000}{0.000} & ${0.938 \pm 0.017}$ & $\mathbf{0.945 \pm 0.004}$ & $\mathbf{0.916 \pm 0.008}$ \\
	GR-M$^-$ & ${0.500 \pm 0.000}$ & ${0.500 \pm 0.000}$ & ${0.500 \pm 0.000}$ & ${0.559 \pm 0.007}$ \\
\bottomrule
\end{tabular}}
	\subfloat[{\large Link Prediction (LP)}]{
\label{tab:link_ablation}
\begin{tabular}{lccccc}\\ \toprule
	\textbf{Models} & \textbf{\grid-T} & \textbf{\communities-T} & \textbf{\grid} & \textbf{\communities} & \textbf{\ppi}\\
	\midrule
	GR-A     & $\mathbf{0.945 \pm 0.021}$ & ${0.990 \pm 0.005}$ & $\mathbf{0.956 \pm 0.014}$ & ${0.991 \pm 0.003}$ & ${0.810 \pm 0.002}$ \\ 
	GR-M     & ${0.940 \pm 0.018}$ & $\mathbf{0.994 \pm 0.003}$  & ${0.931 \pm 0.020}$ & $\mathbf{0.993 \pm 0.003}$ & $\mathbf{0.830 \pm 0.004}$  \\
	GR-M$^-$ & ${0.542 \pm 0.071}$ & ${0.888 \pm 0.046}$ & ${0.500 \pm 0.000}$ & ${0.500 \pm 0.000}$ & ${0.519 \pm 0.026}$ \\
	\bottomrule
\end{tabular}}
}
	\caption{ROC AUC in PNC and LP for ablation study. (GR stands for \name.)}
\label{tab:ablation}
\end{table*}

\addtolength{\tabcolsep}{4pt}

\begin{figure*}[t]
    \centering
    \subfloat[Random Walk Length, $l_w$]{
    \label{fig:walklength}
    \includegraphics[width=\figwidth]{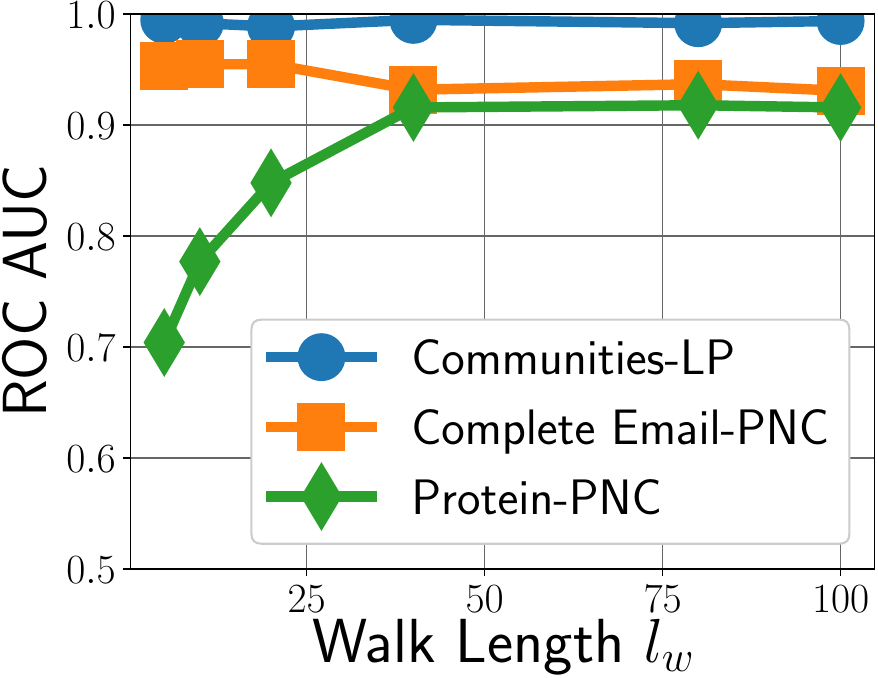}}
    \hfill
    \subfloat[Number of Walks, $n_w$]{
    \label{fig:numwalks}
    \includegraphics[width=\figwidth]{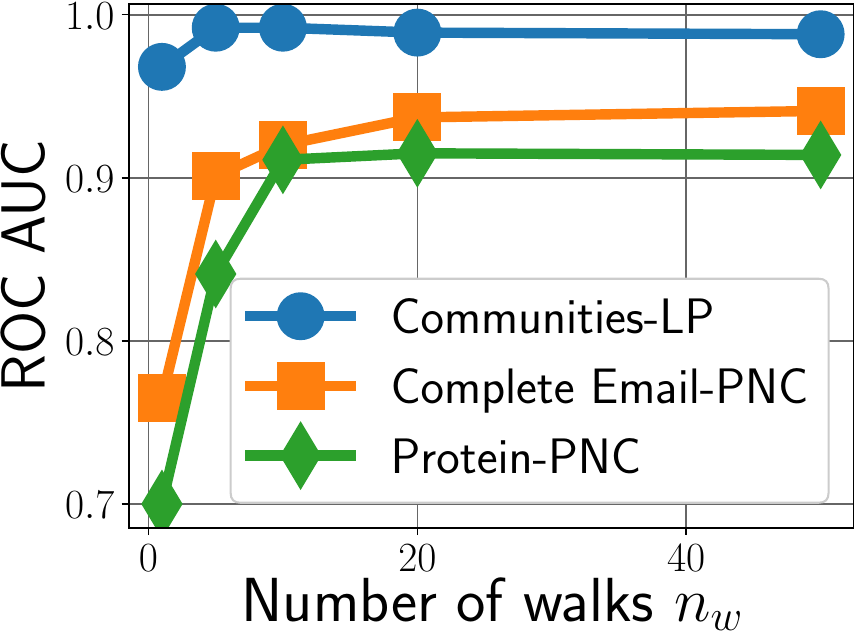}}\hfill
    \subfloat[PNC (Email-Complete)]{
    \label{fig:anchor_pnc}
    \includegraphics[width=\figwidth]{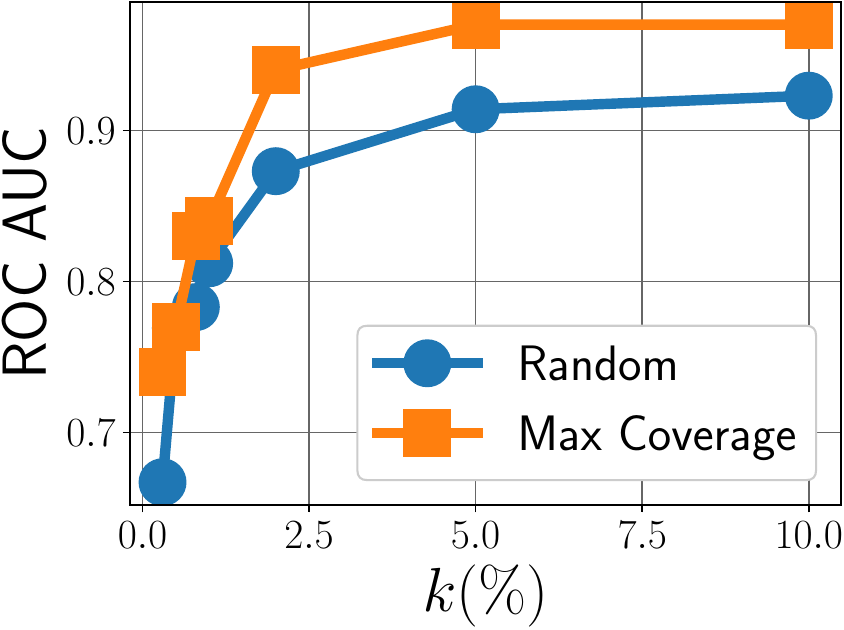}}\hfill
    \subfloat[LP (Communities)]{
    \label{fig:anchor_lp}
    \includegraphics[width=\figwidth]{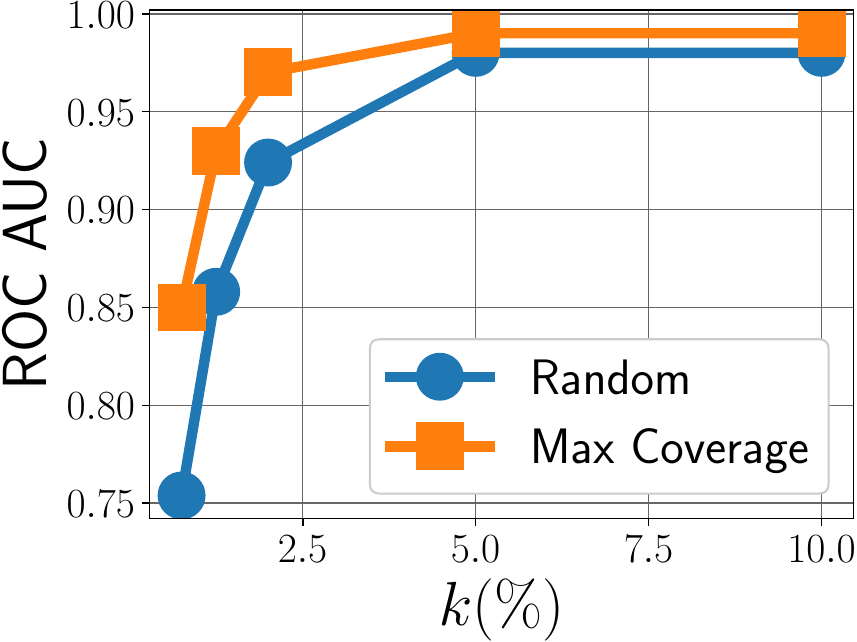}}
	\caption{Impact of (a) walk length (b) number of walks and (c-d) number of anchors for PNC and LP, on accuracy of \name.}
    \label{fig:parameters}
\end{figure*}

\subsection{Ablation Study: Mean Pool versus Attention}
\label{sec:ablation}

In the first two rows of
Table~\ref{tab:ablation}, we compare the performance of \name
with Attention (\name-A) and Mean Pool (\name-M) aggregation
functions. As clearly evident, the performance is comparable.
In mean pool, the message from each anchor is weighted equally,
while when an attention layer is used, \name learns an additional
importance weight for each anchor to aggregate messages. The 
comparable performance of Mean Pool with Attention shows that
similarity encoded in the message from each anchor is enough to
learn meaningful embeddings; the marginal gain from an additional
attention layer is minimal. To further substantiate this claim, we
alter Eq.~\ref{eqn:msg_fn} to
$\mathcal{F}\left(v, a,\ch^l_v, \ch^l_a\right) = \ch^l_v \parallel \ch^l_a$;
i.e, contributions of all anchors are made equal. The variant
\name-M$^-$ presents the performance with this modification; as
evident, there is a massive drop in accuracy. We also evaluate
\name using a similarity function variant (in App.~F).

\subsection{Adversarial Attacks}
\label{sec:adversarial}

We assume the standard \emph{black-box} adversarial setup where the attacker
has knowledge of only the graph and can modify it through addition or
deletion of edges~\cite{adv1,adv2}.  Let $G=(\CV,\CE)$ be the test graph which
has a \emph{colluding} subset of nodes $\CC\subseteq\CV$.
The nodes in $\CC$ can add as many edges as needed among them
so that predictions on $\CC$ are inaccurate. 
%
%
%
For PNC, we randomly sample $10\%$ nodes from the unseen test graph, and form a
clique among these colluding nodes. For LP, we randomly sample $10\%$ of the node
pairs from the unseen test graph such that they do not have an edge between them.
From this set, we select the top-$2\%$ of the highest degree nodes and connnect them
to the remaining colluding nodes through an edge.
This makes the diameter of the colluding group at most $2$. 

We perform prediction using pre-trained models of both \pgnn and \name on the
test graph and measure the ROC AUC on the colluding group. This process is
repeated with $5$ different samples of colluding groups and the mean accuracy
is reported. If the model is robust, despite the collusion, its prediction
accuracy should not suffer.

\begin{table}[t]
	\centering
	\scalebox{0.70}{
	\label{tab:optimal}
	\begin{tabular}{lcccccccc}
	\toprule
		\multirow{2}{*}{\textbf{Dataset}} & \multirow{2}{*}{\textbf{Task}} &\multicolumn{3}{c}{\bf \pgnn} & & \multicolumn{3}{c}{\bf \name} \\
		& &\textbf{Bf} & \textbf{Af} & $\mathbf{\Delta}$ & & \textbf{Bf} & \textbf{Af} & $\mathbf{\Delta}$\\
	\midrule
		Communities & PNC & $0.92$ & $0.82$ & $-0.10$ & & $1.00$ & $0.98$ & $\mathbf{-0.02}$ \\
		Communities & LP  & $1.00$ & $0.89$ & $-0.11$ & & $1.00$ & $1.00$ & $\mathbf{-0.00}$ \\
	\bottomrule
	\end{tabular}}
	\caption{Robustness to adversarial attacks. (\emph{Bf} and \emph{Af} indicate ROC AUC before and after collusion respectively, while $\Delta$ denotes the change in accuracy due to collusion.)}
	\label{tab:adversarial}
\end{table}

As visible in Table~\ref{tab:adversarial}, the impact on the accuracy of \name
is minimal. On the other hand, \pgnn receives more than $10\%$ relative drop in
the performance. This result highlights one more advantage of reachability
estimations. Since \name incorporates all paths in its position-aware
embeddings, causing a significant perturbation in position with respect to all
paths in the network is difficult. In case of \pgnn, perturbing shortest paths
is relatively easier and hence there is a higher impact on the performance.

\subsection{Impact of Parameters}
\label{eq:parameters}

\noindent
\textbf{Random Walk Length, \boldmath$l_w$:} Fig.~\ref{fig:walklength} presents
the result on three datasets covering both PNC and LP. Results on remaining
datasets are provided in App.~F. On \communities and \emailc,
 the impact of $l_w$ is minimal. However, in \protein, the
accuracy saturates at around $l_w=40$. This is a direct consequence of the
property that \protein has a significantly larger diameter of $64$ than both
\communities and \emailc (see Table~\ref{tab:data_stat}). Recall from our
discussion in \S\ref{sec:guarantees} that setting $l_w$ to the graph diameter
is recommended for ensuring accurate reachability estimations. The trends in
Fig.~\ref{fig:walklength} substantiate this theoretical result.

\noindent
\textbf{Number of Random Walks, \boldmath$n_w$:} Fig.~\ref{fig:numwalks}
presents the results. As expected, with higher number of walks, the accuracy
initially improves and then saturates. From \S\ref{sec:guarantees}, we know
that $n_w$ should increase with the number of nodes in the graph to ensure
accurate reachability estimations. The trend in Fig.~\ref{fig:numwalks} is
consistent with this result. In App.~F, we present the results
for this experiment on the remaining datasets.

\noindent
\textbf{Number of Anchors and Anchor Selection Strategy:}
Fig. 3(c-d) present the results. In
addition to the default anchor selection strategy, 
we also evaluate the accuracy obtained
when an equal number of anchors are selected randomly. Two key properties
emerge from this experiment. First, as the number of anchors increases, the
accuracy improves till a saturation point. This is expected since with more
anchors we have more reference points to accurately encode node positions.
Second, the proposed anchor selection strategy is clearly better than random
anchor selection. More importantly, the proposed anchor selection strategy
saturates at around $2.5\%$ compared to $5\%$ in random. Recall that the
dimension of the final embeddings, i.e., the final layer, is equal to the
number of anchors. Consequently, this experiment highlights that high quality
embeddings can be obtained within a low-dimensional space. A low dimension is
preferred in various tasks such as indexing and querying of multi-dimensional
points, due to the adverse impacts of
\emph{curse-of-dimensionality}~\cite{hanansamet}.

\section{Conclusions}
\label{sec:con}

\gnn architectures, despite their impressive success in learning node
embeddings, suffer on predictive tasks that rely on positional information.
Though \pgnn recognized this need, due to reliance on only shortest paths, the
position information captured by \pgnn is not holistic. \name addresses this
limitation and builds a different positional model based on reachability
estimations to anchors computed strategically through fixed-length random walks.
\name achieves a relative improvement of up to $40\%$ over other architectures,
while also being robust to adversarial attacks.

\clearpage
{\small
\bibliographystyle{named}
\bibliography{biblio}

\begin{thebibliography}{}

\bibitem[\protect\citeauthoryear{Borgwardt \bgroup \em et al.\egroup
  }{2005}]{borgwardt2005protein}
Karsten~M Borgwardt, Cheng~Soon Ong, Stefan Sch{\"o}nauer, SVN Vishwanathan,
  Alex~J Smola, and Hans-Peter Kriegel.
\newblock Protein function prediction via graph kernels.
\newblock {\em Bioinformatics}, 21:i47--i56, 2005.

\bibitem[\protect\citeauthoryear{Brin and Page}{1998}]{pagerank}
Sergey Brin and Lawrence Page.
\newblock The anatomy of a large-scale hypertextual web search engine.
\newblock {\em Computer Networks}, 30:107--117, 1998.

\bibitem[\protect\citeauthoryear{Chang \bgroup \em et al.\egroup }{2020}]{adv2}
Heng Chang, Yu~Rong, Tingyang Xu, Wenbing Huang, Honglei Zhang, Peng Cui, Wenwu
  Zhu, and Junzhou Huang.
\newblock A restricted black-box adversarial framework towards attacking graph
  embedding models.
\newblock In {\em AAAI}, pages 3389--3396, 2020.

\bibitem[\protect\citeauthoryear{Cormen \bgroup \em et al.\egroup
  }{2009}]{cormen}
Thomas~H. Cormen, Charles~E. Leiserson, Ronald~L. Rivest, and Clifford Stein.
\newblock {\em Introduction to Algorithms,}.
\newblock The MIT Press, 3rd edition, 2009.

\bibitem[\protect\citeauthoryear{Giles \bgroup \em et al.\egroup
  }{1998}]{giles1998citeseer}
C~Lee Giles, Kurt~D Bollacker, and Steve Lawrence.
\newblock Citeseer: An automatic citation indexing system.
\newblock In {\em Proceedings of the third ACM conference on Digital
  libraries}, pages 89--98, 1998.

\bibitem[\protect\citeauthoryear{Goyal \bgroup \em et al.\egroup
  }{2020}]{graphgen}
Nikhil Goyal, Harsh~Vardhan Jain, and Sayan Ranu.
\newblock Graphgen: A scalable approach to domain-agnostic labeled graph
  generation.
\newblock In {\em Proceedings of The Web Conference 2020}, pages 1253--1263,
  2020.

\bibitem[\protect\citeauthoryear{Grover and
  Leskovec}{2016}]{grover2016node2vec}
Aditya Grover and Jure Leskovec.
\newblock node2vec: Scalable feature learning for networks.
\newblock In {\em ACM SIGKDD}, pages 855--864, 2016.

\bibitem[\protect\citeauthoryear{Gulyás \bgroup \em et al.\egroup
  }{2018}]{sp1}
András Gulyás, Zalán Heszberger, József Bíró, János Tapolcai, Attila
  Csoma, István Pelle, Attila Kőrösi, Dávid Klajbár, Valentina Halasi,
  Gábor Rétvári, and Márton Novák.
\newblock A dataset on human navigation strategies in foreign networked
  systems.
\newblock {\em Scientific Data}, 5, 03 2018.

\bibitem[\protect\citeauthoryear{Hamilton \bgroup \em et al.\egroup
  }{2017}]{hamilton2017graphsage}
Will Hamilton, Zhitao Ying, and Jure Leskovec.
\newblock Inductive representation learning on large graphs.
\newblock In {\em NIPS 2017}, pages 1024--1034, 2017.

\bibitem[\protect\citeauthoryear{Kingma and Ba}{2015}]{kingma2015adam}
Diederik~P. Kingma and Jimmy Ba.
\newblock Adam: {A} method for stochastic optimization.
\newblock In {\em 3rd ICLR}, 2015.

\bibitem[\protect\citeauthoryear{Kipf and Welling}{2017}]{kipf2017iclr}
Thomas~N. Kipf and Max Welling.
\newblock Semi-supervised classification with graph convolutional networks.
\newblock In {\em 5th ICLR}, 2017.

\bibitem[\protect\citeauthoryear{Leskovec \bgroup \em et al.\egroup
  }{2007}]{leskovec2007graph}
Jure Leskovec, Jon Kleinberg, and Christos Faloutsos.
\newblock Graph evolution: Densification and shrinking diameters.
\newblock {\em ACM TKDD}, 2007.

\bibitem[\protect\citeauthoryear{Li \bgroup \em et al.\egroup }{2020}]{adv1}
Jia Li, Honglei Zhang, Zhichao Han, Yu~Rong, Hong Cheng, and Junzhou Huang.
\newblock Adversarial attack on community detection by hiding individuals.
\newblock In {\em WWW 2020}, page 917–927, 2020.

\bibitem[\protect\citeauthoryear{London}{2017}]{sp3}
Transport London.
\newblock {Rolling Origin and Destination Survey}, 2017.

\bibitem[\protect\citeauthoryear{Manchanda \bgroup \em et al.\egroup
  }{2020}]{gcomb}
Sahil Manchanda, Akash Mittal, Anuj Dhawan, Sourav Medya, Sayan Ranu, and Ambuj
  Singh.
\newblock {GCOMB}: Learning budget-constrained combinatorial algorithms over
  billion-sized graphs.
\newblock {\em Advances in Neural Information Processing Systems}, 33, 2020.

\bibitem[\protect\citeauthoryear{McCallum \bgroup \em et al.\egroup
  }{2000}]{mccallum2000cora}
Andrew~Kachites McCallum, Kamal Nigam, Jason Rennie, and Kristie Seymore.
\newblock Automating the construction of internet portals with machine
  learning.
\newblock {\em Information Retrieval}, 3(2):127--163, 2000.

\bibitem[\protect\citeauthoryear{Nemhauser \bgroup \em et al.\egroup
  }{1978}]{submodular}
George~Lann Nemhauser, Laurence~Alexander Wolsey, and Marshall~Latham Fisher.
\newblock An analysis of approximations for maximizing submodular set
  functions-{I}.
\newblock {\em Mathematical Programming}, 14(1):265--294, 1978.

\bibitem[\protect\citeauthoryear{Pan \bgroup \em et al.\egroup }{2004}]{rwr}
Jia-Yu Pan, Hyung-Jeong Yang, Christos Faloutsos, and Pinar Duygulu.
\newblock Automatic multimedia cross-modal correlation discovery.
\newblock In {\em ACM SIGKDD}, pages 653--658, 2004.

\bibitem[\protect\citeauthoryear{Perozzi \bgroup \em et al.\egroup
  }{2014}]{perozzi2014deepwalk}
Bryan Perozzi, Rami Al-Rfou, and Steven Skiena.
\newblock {DeepWalk: Online learning of social representations}.
\newblock In {\em ACM SIGKDD}, pages 701--710, 2014.

\bibitem[\protect\citeauthoryear{Samet}{2006}]{hanansamet}
Hanan Samet.
\newblock {\em Foundations of multidimensional and metric data structures}.
\newblock Academic Press, 2006.

\bibitem[\protect\citeauthoryear{TransStat}{2016}]{sp4}
RITA TransStat.
\newblock {Origin and Destination Survey database (DB1B)}, 2016.

\bibitem[\protect\citeauthoryear{Velickovic \bgroup \em et al.\egroup
  }{2018}]{iclr2018gat}
Petar Velickovic, Guillem Cucurull, Arantxa Casanova, Adriana Romero, Pietro
  Li{\`{o}}, and Yoshua Bengio.
\newblock Graph attention networks.
\newblock In {\em 6th {ICLR}}, 2018.

\bibitem[\protect\citeauthoryear{Wadhwa \bgroup \em et al.\egroup
  }{2019}]{arrival}
Sarisht Wadhwa, Anagh Prasad, Sayan Ranu, Amitabha Bagchi, and Srikanta
  Bedathur.
\newblock Efficiently answering regular simple path queries on large labeled
  networks.
\newblock In {\em ICMD}, pages 1463--1480, 2019.

\bibitem[\protect\citeauthoryear{Watts}{1999}]{watts1999networks}
Duncan~J Watts.
\newblock Networks, dynamics, and the small-world phenomenon.
\newblock {\em American Journal of Sociology}, 105(2):493--527, 1999.

\bibitem[\protect\citeauthoryear{West and Leskovec}{2012}]{sp2}
Robert West and Jure Leskovec.
\newblock Human wayfinding in information networks.
\newblock In {\em WWW}, page 619–628, 2012.

\bibitem[\protect\citeauthoryear{Xu \bgroup \em et al.\egroup
  }{2019}]{xu2019gin}
Keyulu Xu, Weihua Hu, Jure Leskovec, and Stefanie Jegelka.
\newblock How powerful are graph neural networks?
\newblock In {\em 7th {ICLR}}, 2019.

\bibitem[\protect\citeauthoryear{Xu \bgroup \em et al.\egroup }{2020}]{gil}
Chunyan Xu, Zhen Cui, Xiaobin Hong, Tong Zhang, Jian Yang, and Wei Liu.
\newblock Graph inference learning for semi-supervised classification.
\newblock In {\em 8th ICLR}, 2020.

\bibitem[\protect\citeauthoryear{You \bgroup \em et al.\egroup
  }{2019}]{you2019pgnn}
Jiaxuan You, Rex Ying, and Jure Leskovec.
\newblock Position-aware graph neural networks.
\newblock In {\em ICML}, pages 7134--7143, 2019.

\bibitem[\protect\citeauthoryear{Zhu \bgroup \em et al.\egroup
  }{2020}]{homophily}
Jiong Zhu, Yujun Yan, Lingxiao Zhao, Mark Heimann, Leman Akoglu, and Danai
  Koutra.
\newblock Beyond homophily in graph neural networks: Current limitations and
  effective designs.
\newblock In {\em NeurIPS}, 2020.

\bibitem[\protect\citeauthoryear{Zitnik and
  Leskovec}{2017}]{zitnik2017predicting}
Marinka Zitnik and Jure Leskovec.
\newblock Predicting multicellular function through multi-layer tissue
  networks.
\newblock {\em Bioinformatics}, 33(14):i190--i198, 2017.

\end{thebibliography}
}

\clearpage
\section*{Appendix}
\label{sec:paper_appendix}

\appendix
\renewcommand{\thesubsection}{\Alph{subsection}}

\subsection{Order-weighted Similarity}
\label{app:osimilarity}

In Eq.~\ref{eq:sim}, the order in which
nodes occur in the random walk does not affect the similarity function.
To incorporate this
sequential aspect, we propose the use of \emph{harmonic} weighting. Formally,
let $o(v_i,v_j,k)$ denote the step count at which $v_j$ is visited in the
$k^{th}$ random walk originating from node $v_i$; if $v_j$ was not visited,
then $o(v_i,v_j,k)=\infty$. The \emph{order-weighted} similarity is defined
as:
{\small
\begin{align}
\label{eq:osim}
	s_o(v_i,v_j) = \sum^{n_w}_{k=1} \frac{1}{o(v_i,v_j,k)}
\end{align}
}

Please refer to App.~\ref{app:performance_osim} for empirical evaluation of
the model with order-weighted similarity function.

\subsection{NP-hardness of Anchor Selection}
\label{app:nphard}

\begin{thm}
	Anchor selection, which is performed on the bipartite graph formed from the reachability set, is NP-hard.
\end{thm}

\textsc{Proof.} We reduce the \textit{Maximum Coverage problem (MCP)} to the problem of anchor selection.

\begin{defn}[Maximum Coverage]
Given a collection of subsets $\mathbb{S}=\{S_{1},\cdots,S_{m}\}$ from a universal set of items $U=\{t_{1},\cdots,t_{n}\}$ and budget $k$, choose at most $k$ subsets $\mathbb{\mathcal{T}^*}\subseteq \mathbb{S}$ such that the coverage of items $\cup_{\forall S_i\in \mathbb{\mathcal{T}^*}} S_i$ is maximized.
\end{defn}

MCP is known to be NP-hard \cite{cormen}.

Given an arbitrary instance of MCP, we construct a bipartite graph $\mathcal{B}=(\CV_1,\CV_2,\CE_B)$, where we have a node $u_i\in\CV_1$ corresponding to each subset $S_i\in\mathbb{S}$, a node $v_j\in\CV_2$ corresponding to each item $t_j\in U$ and an edge $e=(u_i,v_j)\in\CE_B$ if $S_i$ contains item $t_j$. With this construction, it is easy to see that $\mathcal{A}^*\subseteq\CV_1,\;|\mathcal{A}^*|=k$ maximizes reachability, i.e., $\mid\rho(\mathcal{A}^*)\mid$, if and only if selecting the subsets corresponding to the nodes in $\mathcal{A}^*$ maximizes coverage of items from $U$.
$\square$

\subsection{Monotonicity and Submodularity}
\label{app:submodular} 

\noindent
\textbf{Monotonicity:} The function $f(\CA)=|\rho(\CA)|$ is monotone since
adding any node from $u\in\CV_1$ to $\CA$ can only bring in new neighbors from
$\CV_2$. Hence, $\rho(\CA\cup\{u\})\supseteq \rho(\CA)$ .

\noindent
\textbf{Submodularity:} A function $f(S)$ is \emph{submodular} if the
\emph{marginal gain} from adding an element to a set $S$ is at least as high as
the marginal gain from adding it to a superset of $S$. Mathematically, it
satisfies: 
{\small
\begin{align}
	f(S\cup \{o\})-f(S)\geq f(T \cup \{o\})-f(T)
\end{align}
}
for all elements $o$ and all pairs of sets $S \subseteq T$. 

\begin{lem}
	\label{clm:submodular}
	\textit{Reachability maximization is submodular. Specifically, for any given set of nodes $\CA$, $f(\CA)=|\rho(\CA)|$ is submodular.}
\end{lem}
\textsc{Proof by contradiction:} Assume,
{\small
\begin{align}
	\label{eq:modular1}
	f(T\cup \{v\})-f(T) > f(\CA \cup \{v\})-f(\CA)
\end{align}
}
where $\CA$ and $T$ are subsets of $\CV_1$, such that $\CA\subseteq T$, and $v\in\CV_1$ of bipartite graph $\mathcal{B}$. Given that $f(\CA)$ is monotone, Eq.~\ref{eq:modular1} is feasible only if:
{\small
\begin{align}
	\nonumber
	\rho(\{v\}) \setminus \rho(T) &\supseteq \rho(\{v\}) \setminus \rho(\CA)\\
	\text{or, }\qquad \CA &\not\subseteq T
\end{align}
}
which contradicts the assumption that $\CA\subseteq T$.$\hfill\square$\\

\subsection{Anchor Selection Procedure}
\label{app:greedy}

Algo.~\ref{alg:greedy} outlines the pseudocode for the greedy anchor selection 
procedure. We start from an empty anchor set $\CA$ (line 1), and in each
iteration, add the node $u \in \CV_1$ that has the highest (\emph{marginal})
degree in $\bg$ to $\CA$ (lines 3-4). Following this operation, we remove $u$
from $\CV_1$ and all neighbors of $u$ from $\CV_2$ (lines 5-6). This process
repeats for $k$ iterations (line 2).

\begin{algorithm}[t]
	\caption{Greedy Anchor Selection}
	\label{alg:greedy}
	{\scriptsize
	\textbf{Input:} Graph $\bg=(\CV_1,\CV_2,\CE_B)$; number of anchors: $k$ \\
	\textbf{Output:} $\CA$: a set of $k$ anchors (nodes)
	\begin{algorithmic}[1]
		\STATE $\CA \leftarrow \emptyset$
		\WHILE{$|\CA| < k$}
			\STATE $u^* \leftarrow \argmax_{u\in \CV_1} \{|\rho(\CA\cup \{u\})|-|\rho(\CA)|\}$
			\STATE $\CA \leftarrow \CA\cup \{u^*\}$
			\STATE $\CV_2\leftarrow \CV_2 \setminus \rho(u^*)$
			\STATE $\CV_1\leftarrow \CV_1 \setminus \{u^*\}$
		\ENDWHILE
		\STATE \textbf{return} $\CA$
	\end{algorithmic}
	}
\end{algorithm}

\subsection{Experimental Setup}
\label{app:setup}

All the experiments have been performed on an Intel(R) Xeon(R) Silver 4114 CPU
with a clock speed of 2.20GHz. The GPU used was NVIDIA GeForce RTX 2080 Ti (12GB
of FB memory). We use PyTorch 1.4.0 and NetworkX 2.3 on CUDA 10.0. \name is
implemented in Python 3.7.6. The codebase of all other benchmarked models are
obtained from the respective authors.

\subsubsection{Datasets}
\label{app:datasets}

Below we explain the semantics of each of these datasets.\footnote{All the 
datasets have been taken from \url{https://github.com/JiaxuanYou/P-GNN}\label{fn:code_data}}

\begin{itemize}
	\item \emph{\grid} is a synthetic 2D grid graph with $20 \times 20 = 400$
			nodes and no features.
	\item \emph{\communities} is the connected caveman graph, \cite{watts1999networks}.
			It has $20$ communities of $20$ nodes each.
	\item \emph{\ppi} is a protein-protein interaction network containing $24$ graphs,
			\cite{zitnik2017predicting}.
			Each graph on an average has $3000$ nodes and $33000$ edges.
			Each node is characterized with a $50$-dimensional feature vector.
	\item \emph{\emailc} is a real-world communication graph from SNAP
			\cite{leskovec2007graph}.
	\item \emph{\email} dataset is a set of $7$ graphs obtained by dividing \emailc 
			and has $6$ communities. The label of each node denotes which community it
			belongs to.
	\item \emph{\protein} is a real graph from \cite{borgwardt2005protein}. It contains
			$1113$ components. Each component on average contains $39$ nodes and $73$ edges.
			Each node has $29$ features and is labeled with the functional role of the protein.
	\item \emph{\cora} is a standard citation network of machine-learning papers
			with 2.7K documents, 5.4K links and 1433 distinct word vector attributes,
			divided into seven classes~\cite{mccallum2000cora}.
	\item \emph{\citeseer} is another benchmark citation graph with 3.3K documents,
			4.7K links and 3703 distinct word vector attributes, divided into six
			classes~\cite{giles1998citeseer}.
\end{itemize}

\subsubsection{Predictive Tasks}
\label{app:tasks}

\noindent
\textbf{Link Prediction (LP):} Given a pair of nodes in a graph, the
task is to predict whether there exists a link (edge) between them. \\
\noindent
\textbf{Pairwise Node Classification (PNC):} Given two nodes, the task is to
predict whether these nodes belong to the same class label or come from
different labels. \\
\noindent
\textbf{Node Classification (NC):} For each node, the task is to
predict its class/label.

\subsubsection{Setting}
\label{app:setting}

\noindent
\textbf{Transductive learning: } The nodes are assigned a fixed ordering.
Consequently, the model needs to be re-trained if the ordering changes.
Since the ordering is fixed, \emph{one-hot} vectors can be used as unique
identifiers of nodes. We use these one-hot vectors to augment the node
attributes.

\noindent
\textbf{Inductive learning: } Only the node attributes are used since
under this scenario, the model must generalize to unseen nodes.

\subsubsection{Train-Validation-Test Setup}
\label{app:train_test_setup}

For all prediction tasks, the datasets are
individually split in the ratio of 80:10:10 for training, validation and testing,
respectively. In LP, the positive set contains actual links present in the graph.
The negative set is constructed by sampling an equal number of node pairs that
are not linked. A similar strategy is also applied for PNC. In NC,  we randomly
sample train, validation and test nodes with their corresponding labels. We always
ensure that the test data is unseen. When a graph dataset contains multiple graphs,
we divide each component into train, validation and test sets.

Each experiment (train and test) is repeated
$10$ times following which we report the mean \emph{ROC AUC} and the
standard deviation.

\subsubsection{Common Parameters}
\label{app:parameters}

The number of \emph{hidden layers} is set to $2$. The \emph{hidden
embedding dimension} is set to $32$. All models are trained for
$2000$ epochs. The \emph{learning rate} is set to $0.01$ for the
first $200$ epochs and $0.001$ thereafter. The \emph{drop-out}
parameter is set to $0.5$. \emph{Batch size} is kept to 8 for
Protein and PPI, and $1$ for all other datasets. The final
embeddings for PNC and LP tasks are passed through 1-layer MLPs
characterized by $label(v,u)=\sigma\left(\mathbf{z_v}^T\mathbf{z_u}
\right)$ where $\sigma$ is the \textit{sigmoid} activation
function. The final embeddings for NC are passed through \textit{LogSoftmax}
 layer to get the log-probabilities of each class. In both LP and PNC, the input is
a pair of nodes, while in NC it is a single node. The neural
network parameters are tuned using the Adam optimizer \cite{kingma2015adam}.

\subsection{Performance}
\label{app:performance}

\subsubsection{Efficiency}
\label{app:efficiency}

Table~\ref{tab:times} shows the time taken by the two best performing models on the three largest datasets.
We observe that \name is $2.5$ times faster than \pgnn. \pgnn is slower since it samples a new set of anchors in every layer and epoch, which necessitates the need to recompute distances to all anchors. In contrast, \name uses the same set of strategically chosen anchors through all layers. The inference times of both techniques are less than a second and, hence, is not a computational concern.

\begin{table}[h]
\centering
\resizebox{\columnwidth}{!}{
\begin{tabular}{ccrrrr}
\toprule
\multirow{2}{*}{\textbf{Dataset}} & \multirow{2}{*}{\textbf{Task}} & \multicolumn{2}{c}{\textbf{Training Time (in sec)}} & \multicolumn{2}{c}{\textbf{Inference Time (in sec)}} \\ 
 & & \textbf{\pgnn -E} & \textbf{\name} & \textbf{\pgnn -E} & \textbf{\name} \\
\midrule
\cora &  \multirow{3}{*}{LP} & 326 & 111 & 0.01 &  0.02 \\
\citeseer & & 537 & 125 & 0.01 & 0.01 \\
\ppi & & 5901 & 2980 & 0.20 & 0.20 \\
\midrule
\cora & \multirow{3}{*}{PNC} & 405 & 265 & 0.05 & 0.05 \\
\citeseer & & 645 & 381 & 0.07 & 0.06 \\
\protein & & 13552 & 11254 & 0.90 & 0.90 \\
\bottomrule
\end{tabular}}
\caption{Training and Inference time comparison}
\label{tab:times}
\end{table}

\subsubsection{Order-weighted Similarity}
\label{app:performance_osim}

Table~\ref{tab:osim_ablation} presents the performance achieved with order-weighted similarity function. We observe minimal change in accuracies when compared to using frequency counts (Eq.~\ref{eq:sim}).
A random walker is more likely to visit nearby nodes from the source than those located far away. Consequently, the \emph{early} nodes that receive higher weights in order-weighted similarity are often the same ones that are visited repeatedly. Hence, order-weighting is correlated to count frequency.

\begin{table*}[b]
\centering
\resizebox{\textwidth}{!}
{
\subfloat[Pairwise Node Classification]{
\label{tab:pairwise_osim}
\begin{tabular}{lcccc}\\ \toprule
	\textbf{Models} & \textbf{\communities} & \textbf{\email} & \textbf{\emailc} & \textbf{\protein} \\
	\midrule
	\name-OA & \bld{1.000}{0.000} & ${0.937 \pm 0.009}$ & ${0.936 \pm 0.004}$ & ${0.906 \pm 0.004}$ \\ 
	\name-OM & \bld{1.000}{0.000} & $\mathbf{0.949 \pm 0.014}$ & ${0.934 \pm 0.004}$         & ${0.909 \pm 0.006}$ \\ 
\bottomrule
\end{tabular}}
\subfloat[Link Prediction]{
\label{tab:link_osim}
\begin{tabular}{lccccc}\\ \toprule
	\textbf{Models} & \textbf{\grid-T} & \textbf{\communities-T} & \textbf{\grid} &
		\textbf{\communities} & \textbf{\ppi} \\
		\midrule
	\name-OA & ${0.935 \pm 0.025}$ & ${0.990 \pm 0.005}$ & $\mathbf{0.956 \pm 0.014}$ & ${0.992 \pm 0.004}$ & ${0.822 \pm 0.007}$ \\ 
	\name-OM & ${0.940 \pm 0.025}$  & ${0.993 \pm 0.003}$ & ${0.951 \pm 0.017}$ & $\mathbf{0.993 \pm 0.004}$ & ${0.825 \pm 0.003}$ \\
	\bottomrule
\end{tabular}}}
\caption{{ROC AUC in PNC and LP for ablation study. \name-OA and \name-OM denote Mean Pool and Attention aggregation, respectively, with order-weighted reachability estimation (Eq.~\ref{eq:osim}).}}
\label{tab:osim_ablation}
\end{table*}

\begin{table*}[hbt]
\centering
\resizebox{\textwidth}{!}
{
\subfloat[Pairwise Node Classification]{
\label{tab:lw_pnc}
\begin{tabular}{lcccc}\\ \toprule
	$\mathbf{l_w}$ 	& \textbf{\communities} 	& \textbf{\email} 		& \textbf{\emailc} 			& \textbf{\protein} \\ \midrule
	$5$ 			& ${1.000 \pm 0.000}$ 		& ${0.953 \pm 0.011}$ 	& ${0.922 \pm 0.003}$ 		& ${0.704 \pm 0.167}$\\ 
	$10$ 			& ${1.000 \pm 0.000}$ 		& ${0.955 \pm 0.015}$ 	& ${0.940 \pm 0.007}$ 		& ${0.777 \pm 0.168}$\\ 
	$20$ 			& ${1.000 \pm 0.000}$ 		& ${0.955 \pm 0.010}$ 	& ${0.934 \pm 0.007}$ 		& ${0.848 \pm 0.140}$\\ \midrule
	$40$		    & ${1.000 \pm 0.000}$ 		& ${0.922 \pm 0.020}$ 	& ${0.937 \pm 0.008}$ 		& ${0.916 \pm 0.004}$\\ 
	$80$ 			& ${1.000 \pm 0.000}$ 		& ${0.927 \pm 0.011}$ 	& ${0.936 \pm 0.005}$ 		& ${0.918 \pm 0.007}$\\ 
	$100$ 			& ${1.000 \pm 0.000}$ 		& ${0.921 \pm 0.022}$ 	& ${0.943 \pm 0.003}$ 		& ${0.916 \pm 0.004}$\\ \bottomrule
\end{tabular}}
\subfloat[Link Prediction]{
\label{tab:lw_lp}
\begin{tabular}{lccccc}\\ \toprule
	$\mathbf{l_w}$ 		& \textbf{\grid-T} 		& \textbf{\communities-T} 	& \textbf{\grid} 		&\textbf{\communities} 		& \textbf{\ppi}                                      \\ \midrule
	$5$         		& ${0.921 \pm 0.033}$ 	& ${0.989 \pm 0.005}$ 		& ${0.938 \pm 0.021}$ 	& ${0.995 \pm 0.002}$ 		& ${0.815 \pm 0.006}$ \\
	$10$ 				& ${0.922 \pm 0.015}$ 	& ${0.991 \pm 0.003}$ 		& ${0.939 \pm 0.011}$ 	& ${0.992 \pm 0.002}$ 		& ${0.827 \pm 0.005}$ \\
	$20$ 				& ${0.951 \pm 0.014}$ 	& ${0.993 \pm 0.002}$ 		& ${0.938 \pm 0.017}$ 	& ${0.989 \pm 0.002}$ 		& ${0.833 \pm 0.001}$ \\ \midrule
	$40$ 				& ${0.952 \pm 0.013}$ 	& ${0.991 \pm 0.003}$ 		& ${0.948 \pm 0.020}$ 	& ${0.995 \pm 0.001}$ 		& ${0.830 \pm 0.004}$ \\
	$80$  				& ${0.920 \pm 0.033}$ 	& ${0.993 \pm 0.004}$ 		& ${0.940 \pm 0.015}$ 	& ${0.992 \pm 0.004}$ 		& ${0.823 \pm 0.004}$ \\
	$100$ 				& ${0.938 \pm 0.026}$ 	& ${0.992 \pm 0.003}$ 		& ${0.935 \pm 0.029}$ 	& ${0.994 \pm 0.002}$ 		& ${0.821 \pm 0.005}$ \\\bottomrule
\end{tabular}}}
\caption{{Effect of the length of random walk on accuracy of \name.}}
\label{tab:lw}
\end{table*}

\subsubsection{Impact of Length of Random Walk}
\label{app:walklength}

The effect of length of walk $l_w$ is summarized in Table \ref{tab:lw_pnc} and Table \ref{tab:lw_lp}.
We observe that for small datasets such as \communities, \email, \emailc and \ppi, the ROC AUC scores
start saturating when $l_w$ is around $10$, which is approximately the diameter of the datasets.
The accuracy on the \grid dataset saturates for $l_w = 20$ (diameter is $38$) while for \protein it saturates
at $l_w = 40$ (diameter is $64$).

\begin{table*}[hbt]
\centering
\resizebox{\textwidth}{!}
{
\subfloat[Pairwise Node Classification]{
\label{tab:nw_pnc}
\begin{tabular}{lcccc}\\ \toprule
	$\mathbf{n_w}$ 	& \textbf{\communities} 	& \textbf{\email} 		& \textbf{\emailc} 			& \textbf{\protein} \\ \midrule
	$1$ 			& ${0.999 \pm 0.002}$ 		& ${0.882 \pm 0.018}$ 	& ${0.765 \pm 0.014}$ 		& ${0.700 \pm 0.165}$\\ 
	$5$ 			& ${1.000 \pm 0.000}$ 		& ${0.946 \pm 0.007}$ 	& ${0.901 \pm 0.008}$ 		& ${0.841 \pm 0.140}$\\ 
	$10$ 			& ${1.000 \pm 0.000}$ 		& ${0.909 \pm 0.030}$ 	& ${0.920 \pm 0.009}$ 		& ${0.911 \pm 0.004}$\\ 
	$20$ 			& ${1.000 \pm 0.000}$ 		& ${0.950 \pm 0.014}$ 	& ${0.937 \pm 0.003}$ 		& ${0.915 \pm 0.007}$\\ \midrule
	$50$		    & ${1.000 \pm 0.000}$ 		& ${0.949 \pm 0.016}$ 	& ${0.941 \pm 0.003}$ 		& ${0.914 \pm 0.002}$\\
	$75$ 			& ${1.000 \pm 0.000}$ 		& ${0.936 \pm 0.008}$ 	& ${0.941 \pm 0.005}$ 		& ${0.916 \pm 0.003}$\\
	$100$ 			& ${1.000 \pm 0.000}$ 		& ${0.932 \pm 0.021}$ 	& ${0.939 \pm 0.004}$ 		& ${0.920 \pm 0.004}$\\ \midrule
	$200$ 			& ${1.000 \pm 0.000}$ 		& ${0.955 \pm 0.007}$ 	& ${0.940 \pm 0.004}$ 		& ${0.916 \pm 0.003}$\\ 
	$500$ 			& ${1.000 \pm 0.000}$ 		& ${0.947 \pm 0.006}$ 	& ${0.948 \pm 0.004}$ 		& ${0.914 \pm 0.006}$\\ \bottomrule
\end{tabular}}
\subfloat[Link Prediction]{
\label{tab:nw_lp}
\begin{tabular}{lccccc}\\ \toprule
	$\mathbf{n_w}$ 		& \textbf{\grid-T} 		& \textbf{\communities-T} 	& \textbf{\grid} 		&\textbf{\communities} 		& \textbf{\ppi}                                      \\ \midrule
	$1$         		& ${0.739 \pm 0.048}$ 	& ${0.988 \pm 0.006}$ 		& ${0.805 \pm 0.036}$ 	& ${0.968 \pm 0.013}$ 		& ${0.624 \pm 0.028}$ \\
	$5$         		& ${0.920 \pm 0.021}$ 	& ${0.991 \pm 0.004}$ 		& ${0.898 \pm 0.009}$ 	& ${0.992 \pm 0.002}$ 		& ${0.790 \pm 0.006}$ \\
	$10$ 				& ${0.926 \pm 0.023}$ 	& ${0.992 \pm 0.002}$ 		& ${0.941 \pm 0.023}$ 	& ${0.992 \pm 0.002}$ 		& ${0.812 \pm 0.003}$ \\ 
	$20$ 				& ${0.935 \pm 0.012}$ 	& ${0.993 \pm 0.006}$ 		& ${0.932 \pm 0.019}$ 	& ${0.989 \pm 0.002}$ 		& ${0.825 \pm 0.005}$ \\ \midrule
	$50$ 				& ${0.924 \pm 0.026}$ 	& ${0.990 \pm 0.003}$ 		& ${0.946 \pm 0.018}$ 	& ${0.988 \pm 0.002}$ 		& ${0.834 \pm 0.002}$ \\
	$75$  				& ${0.932 \pm 0.013}$ 	& ${0.990 \pm 0.004}$ 		& ${0.961 \pm 0.017}$ 	& ${0.996 \pm 0.001}$ 		& ${0.832 \pm 0.005}$ \\ 
	$100$ 				& ${0.949 \pm 0.010}$ 	& ${0.992 \pm 0.002}$ 		& ${0.955 \pm 0.016}$ 	& ${0.992 \pm 0.004}$ 		& ${0.829 \pm 0.004}$ \\ \midrule
	$200$ 				& ${0.926 \pm 0.018}$ 	& ${0.993 \pm 0.002}$ 		& ${0.925 \pm 0.018}$ 	& ${0.990 \pm 0.002}$ 		& ${0.824 \pm 0.003}$ \\
	$500$ 				& ${0.938 \pm 0.005}$ 	& ${0.994 \pm 0.001}$ 		& ${0.938 \pm 0.018}$ 	& ${0.992 \pm 0.004}$ 		& ${0.818 \pm 0.006}$ \\ \bottomrule	
\end{tabular}}}
\caption{{Effect of the number of walks on accuracy of \name.}}
\label{tab:nw}
\end{table*}

\subsubsection{Impact of Number of Random Walks}
\label{app:numwalks}

As discussed earlier, theoretically, the number of random walks conducted is
proportional to the number of nodes of the graph. We observe from results shown
in Table \ref{tab:nw_pnc} and Table \ref{tab:nw_lp} that saturation point for
all the datasets was achieved for a small number of random walks.

\end{document}